\documentclass[pra,aps,twocolumn,superscriptaddress,showpacs,showkeys,nofootinbib]{revtex4-1}
\usepackage[latin1]{inputenc}
\usepackage{amsmath,amsthm,amssymb,graphicx,subfigure,xcolor}
\allowdisplaybreaks[4]
\usepackage{mathrsfs}
\usepackage{graphics,float}
\usepackage{epstopdf}
\usepackage{color}
\usepackage{xcolor}
\usepackage[unicode=true]{hyperref}
\usepackage{booktabs}
\usepackage{tabularx}
\usepackage{tabu}
\usepackage{multirow}
\hypersetup{
     colorlinks=true,       		
     linkcolor=red,          	
     citecolor=blue,            
     urlcolor=blue,           	
 }
\newtheorem{theorem}{Theorem}
\newtheorem*{theorem*}{Theorem}

\newtheorem*{corollary*}{Corollary}
\newtheorem{lemma}{Lemma}
\newtheorem*{lemma*}{Lemma}

\newtheorem*{proposition*}{Proposition}
\theoremstyle{definition}

\newtheorem*{definition*}{Definition}
\theoremstyle{remark}

\newtheorem*{remark*}{Remark}

\newtheorem*{example*}{Example}
\newtheorem{counterexample}{Counterexample}
\def\ba{\begin{array}}
\def\ea{\end{array}}
\def\be{\begin{equation}}
\def\ee{\end{equation}}

\newcommand{\ket}[1]{|#1\rangle}

\newcommand{\Tr}{\rm Tr}
\begin{document}
\title{Generalized product-form monogamy relations in multi-qubit systems}

\author{Wen Zhou}
\email{2230501027@cnu.edu.cn}
\affiliation{School of Mathematical Sciences, Capital Normal University, Beijing 100048, China}
\author{Zhong-Xi Shen}
\email{18738951378@163.com}
\affiliation{School of Mathematics and Physics, Nanyang Institute of Technology, Nanyang, Henan 473004, China}
\author{Hong-Xing Wu}
\email{2240501014@cnu.edu.cn}
\affiliation{School of Mathematical Sciences, Capital Normal University, Beijing 100048, China}
\affiliation{School of Mathematics and Computational Science, Shangrao Normal University, Shangrao 334001, China}
\author{Zhi-Xi Wang}
\email{wangzhx@cnu.edu.cn}
\affiliation{School of Mathematical Sciences, Capital Normal University, Beijing 100048, China}
\author{Shao-Ming Fei}
\email{feishm@cnu.edu.cn}
\affiliation{School of Mathematical Sciences, Capital Normal University, Beijing 100048, China}

\begin{abstract}
Monogamy of entanglement essentially characterizes the entanglement distributions among the subsystems. Generally it is given by summation-form monogamy inequalities. In this paper, we present the product-form monogamy inequalities satisfied by the $\nu$-th ($\nu\geq2$) power of the concurrence. We show that they are tighter than the existing ones by detailed example. We then establish tighter product-form monogamy inequalities based on the negativity. We show that they are valid even for high dimensional states to which the well-known CKW inequality is violated.\\

\end{abstract}

\maketitle

\section{Introduction}
Quantum entanglement~\cite{MAN,RPMK,KSS,HPB,JIV,CYSG} is a fundamental issue of quantum mechanics. It plays a pivotal role in distinguishing the quantum from the classical world.
An important feature of quantum entanglement is the monogamy, which limits the sharability of quantum entanglement among many-body quantum systems. For an entanglement measure $\mathcal{E}$ of bipartite states, Coffman, Kundu, and Wootters (CKW)~\cite{CKW2000} first characterized the monogamy of entanglement (MOE) for the three-qubit state mathematically:
\begin{equation*}
\mathcal{E}(\varrho_{A|BC})\geq \mathcal{E}(\varrho_{AB})+\mathcal{E}(\varrho_{AC}),
\end{equation*}
where $\varrho_{AB}={\rm Tr}_C(\varrho_{ABC})$, $\varrho_{AC}={\rm Tr}_B(\varrho_{ABC})$ and $\mathcal{E}(\varrho_{A|BC})$ stands for the entanglement under bipartition $A$ and $BC$. Osborne and Verstraete demonstrated that the squared concurrence satisfies the monogamy inequality for any $N$-qubit systems ~\cite{Osborne220503}. Monogamy relations have also been extensively explored for various quantum correlations, including quantum discord~\cite{asga,yknz}, quantum steering~\cite{mdam,czsc} and Bell nonlocality~\cite{vsng}.

The generalized summation-form monogamy relations in terms of effective entanglement measurements have been investigated in \cite{Zhu0243042014,Jin0323362008,Shen274,Shen127}.
Different from the original monogamy inequality, the authors of Ref.~\cite{Zhang1553042020} introduced a product-form monogamy inequality. Subsequently, Zhang {\it et. al} explored the product-form monogamy relations for multipartite entanglement, specifically in terms of the $\nu$-th ($\nu\geq2$) power of concurrence and negativity~\cite{Zhang21500222021}.

In this paper, we present a tighter monogamy relations in the product-form for concurrence and negativity.  Regarding the relations among the summation-form and product-form monogamy relations, we show that the product-form monogamy relation possesses a stricter lower bound. Furthermore, it is shown that the newly proposed product-form monogamy relations are more efficient in dealing with the counterexamples raised by the CKW monogamy inequality in higher-dimensional systems.

\section{Enhanced product-form monogamy relations for concurrence}\label{s1}
For a bipartite pure state $|\psi\rangle_{AB}$ in finite dimensional Hilbert space $H_A\otimes H_B$, the concurrence is presented by~\cite{U032307,RBCHM}
\begin{eqnarray}\label{defconcurrence}
\mathcal{C}(|\psi\rangle_{AB})=\sqrt{{2\left[1-\mathrm{Tr}(\varrho_A^2)\right]}},
\end{eqnarray}
where $\varrho_A=\mathrm{Tr}_B(|\psi\rangle_{AB}\langle\psi|)$ denotes the reduced density matrix. The concurrence of a 2-qubit mixed state $\varrho$ is expressed as \cite{W2245}
\begin{eqnarray*}
\mathcal{C}(\varrho)=\max\{\vartheta_1-\vartheta_2-\vartheta_3-\vartheta_4,0\},
\end{eqnarray*}
where $\vartheta_1\geq\vartheta_2\geq\vartheta_3\geq\vartheta_4$ are the eigenvalues of the matrix $\sqrt{\sqrt{\varrho}\widetilde{\varrho}\sqrt{\varrho}}$, with $\widetilde{\varrho}=(\varrho_y\otimes\varrho_y) \varrho^\ast (\varrho_y\otimes\varrho_y)$, $\varrho^\ast$ being the complex conjugation of $\varrho$ and $\varrho_y$ the standard Pauli matrix. It has been shown that for a three-qubit state $\varrho_{ABC}$,
\begin{eqnarray}\label{sum0}
\mathcal{C}^2(\varrho_{A|BC})\geq \mathcal{C}^2(\varrho_{AB})+ \mathcal{C}^2(\varrho_{AC}),
\end{eqnarray}
which implies that sum of the entanglement shared between $AB$ and $AC$ is restricted by the entanglement shared between $A$ and $BC$.

For any $N$-qubit state $\varrho_{AB_1 \cdots B_{N-1}}$ in $H_A\otimes H_{B_1}\otimes \cdots \otimes H_{B_{N-1}}$, the concurrence $ \mathcal{C}(\varrho_{A|B_1 \cdots B_{N-1}})$ of the state $\varrho_{AB_1 \cdots B_{N-1}}$ under the bipartition $A$ and $B_1,B_2,\cdots,B_{N-1}$ satisfies~\cite{Zhu0243042014}
\begin{equation}\label{sum1}
\mathcal{C}^{\nu}(\varrho_{A|B_1 \cdots B_{N-1}})\geq  \mathcal{C}^{\nu}(\varrho_{AB_1})+\cdots+ \mathcal{C}^{\nu}(\varrho_{AB_{N-1}})
\end{equation}
for all $\nu\geq2$, where $\varrho_{AB_i}$ represents the two-qubit reduced density matrices of subsystems $AB_{i}$, $i=1,2,\cdots,N-1$.

Besides the summation-form monogamy inequality (\ref{sum0}), one has also product-form monogamy relations satisfied by the squared concurrence \cite{Zhang1553042020},
\begin{equation}\label{product1}
\mathcal{C}^2(\varrho_{A|BC})\geq 2\big(\mathcal
{C}(\varrho_{AB})^2\mathcal{C}(\varrho_{AC})^2+\frac{\kappa_{ABC}^2}{4}\big)^{\frac{1}{2}},
\end{equation}
where $\kappa_{ABC}$ is the residual entanglement \cite{CS2007},
\begin{equation*}
\kappa_{ABC}=\mathcal{C}^2(\varrho_{A|BC})-\big(\mathcal{C}^2(\varrho_{AB})
+\mathcal{C}^2(\varrho_{AC})\big).
\end{equation*}
Later, Zhang {\it  et. al} investigated the product-form monogamy relations in terms of the $\nu$-th ($\nu\geq2$) power of concurrence~\cite{Zhang21500222021},
\begin{equation}\label{product1.1}
\mathcal{C}^\nu(\varrho_{A|BC})\geq
\big(4\mathcal{C}(\varrho_{AB})^2\mathcal{C}(\varrho_{AC})^2
+\kappa_{ABC}^2\big)^{\frac{\nu}{4}}.
\end{equation}

In what follows, we derive product-form monogamy inequalities which are tighter than the inequality derived in Ref.~\cite{Zhang21500222021}. We introduce two lemmas to study the product-form monogamy relations of entanglement concurrence in multi-qubit systems. For convenience, we represent by $\mathcal{C}_{AB_j}=\mathcal{C}(\varrho_{AB_j})$ for $j=1,2,\cdots,N-1$, and $\mathcal{C}_{A|B_1B_2\cdots B_{N-1}}=\mathcal{C}(\varrho_{A|B_1 B_2\cdots B_{N-1}})$.

\begin{lemma}\label{productL1}
For any three-qubit pure state $|\psi\rangle_{ABC}\in H_A\otimes H_B\otimes H_C$, we have
\begin{equation}\label{product5}
\begin{aligned}
\mathcal{C}^{\nu}_{A|BC}\geq \big[4(\mathcal{C}^2_{AB}+\frac{\kappa_{ABC}}{2})(\mathcal{C}^2_{AC}
+\frac{\kappa_{ABC}}{2})\big]^{\frac{\nu}{4}}
\end{aligned}
\end{equation}
for $\nu\geq2$.

\begin{proof}
For a three-qubit pure state $|\psi\rangle_{ABC}$, the concurrence $\mathcal {C}_{AB}$ satisfies \cite{CKW2000}
\begin{equation}\label{product2}
\mathcal{C}^2_{AB}=\Tr(\varrho_{AB}\tilde{\varrho}_{AB})-2\vartheta_1\vartheta_2,
\end{equation}
where $\varrho_{AB}$ is the reduced density matrix and $\vartheta_1, \vartheta_2$ are the square roots of two non zero eigenvalues of $\varrho_{AB}\tilde\varrho_{AB}$. Since, $\vartheta_1\vartheta_2=\frac{\kappa_{ABC}}{4}$, Eq.~(\ref{product2}) becomes
\begin{equation}\label{product3}
\mathcal{C}^2_{AB}=\Tr(\varrho_{AB}\tilde{\varrho}_{AB})-\frac{\kappa_{ABC}}{2}.
\end{equation}
Similarly for $AC$, we have
\begin{equation}\label{product4}
\mathcal{C}^2_{AC}=\Tr(\varrho_{AC}\tilde{\varrho}_{AC})-\frac{\kappa_{ABC}}{2}.
\end{equation}

On other hand, the concurrence $\mathcal{C}_{A|BC}$  between partition $A$ and $BC$ has the form, $\Tr(\varrho_{AB}\tilde{\varrho}_{AB})+\Tr(\varrho_{AC}\tilde{\varrho}_{AC})=\mathcal{C}^2_{A|BC}$~\cite{CKW2000}.
Using Eq.~(\ref{product3}) and (\ref{product4}), we have
\begin{align*}
\mathcal{C}^2_{A|BC}&=\Tr(\varrho_{AB}\tilde{\varrho}_{AB})+\Tr(\varrho_{AC}\tilde{\varrho}_{AC})\\
&\geq2\big[(\Tr(\varrho_{AB}\tilde{\varrho}_{AB})(\Tr(\varrho_{AC}\tilde{\varrho}_{AC}) \big]^{\frac{1}{2}}\\
&= 2\big[(\mathcal{C}^2_{AB}+\frac{\kappa_{ABC}}{2})(\mathcal{C}^2_{AC}
+\frac{\kappa_{ABC}}{2})\big]^{\frac{1}{2}},
\end{align*}
where  the inequality holds as $\upsilon_{1}^2+\upsilon_{2}^2\geq 2\upsilon_{1}\upsilon_{2}$ for $\upsilon_{1},\upsilon_{2} \geq 0$. Therefore, for $\nu\geq2$ we obtain
\begin{equation*}
\begin{array}{rl}
\mathcal{C}^\nu_{A|BC}=&\big(\mathcal{C}^2_{A|BC}\big)^{\frac{\nu}{2}}\\
&\geq \Big[2\sqrt{(\mathcal{C}^2_{AB}+\frac{\kappa_{ABC}}{2})
(\mathcal{C}^2_{AC}+\frac{\kappa_{ABC}}{2})}\Big]^{\frac{\nu}{2}}\\[1mm]
&=\big[4(\mathcal{C}^2_{AB}+\frac{\kappa_{ABC}}{2})
(\mathcal{C}^2_{AC}+\frac{\kappa_{ABC}}{2})\big]^{\frac{\nu}{4}}.
\end{array}
\end{equation*}
\end{proof}
\end{lemma}

{\noindent\bf Remark 1}. It is obvious that the product-form monogamy inequality (\ref{product5})  is stricter than the inequality (\ref{product5})  proposed in Ref.~\cite{Zhang21500222021}. Now let us take into account the $3$-qubit $W$ state,

$$|W\rangle=\frac{|100\rangle+|010\rangle+|001\rangle}{\sqrt{3}}.$$
One has~\cite{ZXN},
$\mathcal{C}^{2}_{A|BC}=\mathcal{C}^{2}_{AB}+\mathcal{C}^{2}_{AC}$.
Hence, the residual entanglement $\kappa_{ABC} =0$ and the product-form monogamy inequality (\ref{product5}) reduces to
\begin{equation}
\begin{aligned}
\mathcal{C}^{\nu}_{A|BC}\geq \big[4(\mathcal{C}^2_{AB})(\mathcal{C}^2_{AC})\big]^{\frac{\nu}{4}}\nonumber.
\end{aligned}
\end{equation}
Evidently, the inequality (\ref{product1.1}) in Ref.~\cite{Zhang21500222021} is just a specific case of our Lemma \ref{productL1}. When $\nu=2$, Lemma \ref{productL1} reduces to the result (\ref{product1})  presented in Ref.~\cite{Zhang1553042020}.

\begin{lemma}\label{productL2}
For any $N$-qubit pure state $\varrho_{AB_1\cdots B_{N-1}}\in H_{A}\otimes H_{B_1}\otimes \cdots\otimes H_{B_{N-1}}$,
\small\begin{equation}\label{product6}
\begin{aligned}
\big(\prod\limits_{i=1}^{N-1}\mathcal{C}^2_{AB_{i}}\big)^{\frac{1}{N-1}}
\leq\frac{1}{N-1}\sum\limits_{i=1}^{N-1}\mathcal{C}^2_{AB_{i}}
\leq\frac{1}{N-1}\mathcal{C}^2_{A|B_{1}B_{2}\cdots B_{N-1}}.
\end{aligned}
\end{equation}\normalsize

\begin{proof}
Through the utilization of inequality (\ref{sum1}) and the arithmetic-geometric mean inequality, we obtain inequality (\ref{product6}).
\end{proof}
\end{lemma}

Based on Lemmas \ref{productL1} and \ref{productL2}, we have the following product-form monogamy relations given by the $\nu$-th power of concurrence.

\begin{theorem}\label{productT1}
For any $N$-qubit pure quantum state $\varrho_{AB_1\cdots B_{N-1}}\in H_{A}\otimes H_{B_1}\otimes \cdots\otimes H_{B_{N-1}}$, when $\nu\geq2$, one has
\begin{equation}\label{product7}
\begin{array}{rl}
\mathcal{C}_{A|B_1\cdots B_{N-1}}^\nu\geq
&\Big(4(\mathcal{C}^2_{AB_{1}}+\frac{\kappa_{AB_{1}\cdots B_{N-1}}}{2})\\
&\big((N-2)\Big(\prod\limits_{i=1}^{N-2}\mathcal{C}^2_{AB_{i+1}}\Big)^{\frac{1}{N-2}}\\
&+\frac{\kappa_{AB_{1}\cdots B_{N-1}}}{2}\big)\Big)^{\frac{\nu}{4}},
\end{array}
\end{equation}
where $\kappa_{AB_1\cdots B_{N-1}}=\mathcal{C}^2(\varrho_{A|B_1\cdots B_{N-1}})-\big(\mathcal{C}^2(\varrho_{AB_1})+\mathcal{C}^2(\varrho_{AB_2\cdots B_{N-1}})\big).$

\begin{proof}
According to Lemma \ref{productL1}, we get
\begin{equation*}
\begin{array}{rl}
&\mathcal{C}_{A|B_1\cdots B_{N-1}}^\nu\\
\geq&\big[4(\mathcal{C}^2_{AB_{1}}+\frac{\kappa_{AB_{1}\cdots B_{N-1}}}{2})(\mathcal{C}^2_{A|B_{2}\cdots B_{N-1}}+\frac{\kappa_{AB_{1}\cdots B_{N-1}}}{2})\big]^{\frac{\nu}{4}}\\
\geq&\big[4(\mathcal{C}^2_{AB_{1}}+\frac{\kappa_{AB_{1}\cdots B_{N-1}}}{2})(\sum\limits_{i=1}^{N-2}\mathcal{C}^2_{AB_{i+1}}+\frac{\kappa_{AB_{1}\cdots B_{N-1}}}{2})\big]^{\frac{\nu}{4}}\\
\geq&\big[4(\mathcal{C}^2_{AB_{1}}+\frac{\kappa_{AB_{1}\cdots B_{N-1}}}{2})\big((N-2)\Big(\prod\limits_{i=1}^{N-2}\mathcal{C}^2_{AB_{i+1}}\Big)^{\frac{1}{N-2}}\\
&+\frac{\kappa_{AB_{1}\cdots B_{N-1}}}{2}\big)\big]^{\frac{\nu}{4}}.
\end{array}
\end{equation*}
Taking into account the inequalities (\ref{sum1}) and (\ref{product6}), we complete the proof.
\end{proof}
\end{theorem}

Compared with the summation-form monogamy relation (\ref{sum1}) Ref.~\cite{Zhu0243042014}, it is seen that the lower bound of the product-form monogamy relation (\ref{product7}) may be tighter. We consider the following example to illustrate the validity of our product-form monogamy relation of multi-qubit entanglement.

{\noindent\bf Example 1}. Let us consider the following three-qubit state~\cite{Tr39},
\begin{equation}\label{producteg1}
\begin{array}{rl}
|\phi\rangle= p_1e^{i\theta}|000\rangle + p_2|001\rangle +p_3|010\rangle +p_4|100\rangle +p_5 |111\rangle,
\end{array}
\end{equation}
where $p_i>0$, $i=0,1,\cdots,4$, $\sum\limits_{i=0}^{4}p_i^2=1$ and $0 \leq \theta < \pi$.
Let $\theta=0$. One has
$\mathcal{C}^2_{AB}= 4(p_3p_4-p_2p_5)^2$, $\mathcal{C}^2_{AC}= 4(p_2p_4-p_3p_5)^2$ and $ \mathcal{C}^2_{A|BC} = -4(p_4^2-p_5^2+p_5^4+p_4^2(-1+p_1^2+2p_5^2))$. The residual entanglement is given by $\kappa_{ABC} = 4p_5^2(4p_2p_3p_4+p_1^2p_5)^2$.
Setting $p_1=p_5=\frac{1}{5}$, $p_2=\frac{\sqrt{15}}{5}$ and $p_3=p_4=\frac{2}{5}$, we obtain
$\mathcal{C}_{A|BC}=\sqrt{\frac{48}{625}}$, $\mathcal{C}_{AB}=\frac{2(4-\sqrt{15})}{25}$, $\mathcal{C}_{AC}=\frac{2(2\sqrt{5}-2)}{25}$ and $\kappa_{ABC}=\frac{4}{25}(\frac{16\sqrt{15}+1}{125})^{2}$.

\begin{figure}[htbp]
\centering
{\includegraphics[width=6cm,height=4.5cm]{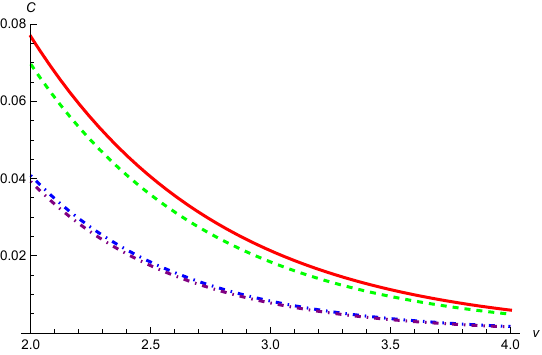}}
\caption{Solid red line denotes $\mathcal{C}^\nu_{A|BC}$ for the state given in Eq.(\ref{producteg1}). The green thick dotted (blue dot dashed thick, purple dot dashed thick) line represents the lower bound from our result (\ref{product5}) (Eq.(\ref{product1.1}) in Ref.~\cite{Zhang21500222021} and Eq.(\ref{sum1})  in Ref.~\cite{Zhu0243042014}, respectively).} \label{Fig1}
\end{figure}

Then we have $\mathcal{C}^{\nu}_{AB}+\mathcal{C}^{\nu}_{AC}
=(\frac{2(4-\sqrt{15})}{25})^{\nu}+(\frac{2(2\sqrt{5}-2)}{25})^{\nu}$
from Eq.(\ref{sum1}) in Ref.~\cite{Zhu0243042014},
$\big(4\mathcal{C}_{AB}^2\mathcal{C}_{AC}^2+\kappa_{ABC}^2\big)^{\frac{\nu}{4}}
=\big(4(\frac{2(4-\sqrt{15})}{25})^2(\frac{2(2\sqrt{5}-2)}{25})^2
+(\frac{4}{25}(\frac{16\sqrt{15}+1}{125})^{2})^2\big)^{\frac{\nu}{4}}$ from Eq.(\ref{product1.1}) in Ref.~\cite{Zhang21500222021} and $\big[4(\mathcal{C}^2_{AB}+\frac{\kappa_{ABC}}{2})(\mathcal{C}^2_{AC}
+\frac{\kappa_{ABC}}{2})\big]^{\frac{\nu}{4}}
=\big[4((\frac{2(4-\sqrt{15})}{25})^2+\frac{2}{25}(\frac{16\sqrt{15}+1}{125})^{2})
((\frac{2(2\sqrt{5}-2)}{25})^2+\frac{2}{25}
(\frac{16\sqrt{15}+1}{125})^{2})\big]^{\frac{\nu}{4}}$
from our result (\ref{product5}). It is evident that our result (\ref{product5}) is superior to the results of Eq.(\ref{sum1}) and Eq.(\ref{product1.1}) presented in Ref.~\cite{Zhu0243042014} and Ref.~\cite{Zhang21500222021}, respectively, see Fig.\ref{Fig1}.

\section{Enhanced product-form monogamy relations for negativity}\label{s2}
Another widely known quantifier of bipartite entanglement is the negativity. For a bipartite state $\varrho_{AB}$ in $H_A\otimes H_B$, the negativity is defined as follows \cite{Vidal0323142002}
\begin{equation*}
 N(\varrho_{AB})=\frac{||\varrho_{AB}^{T_A}||-1}{2},
\end{equation*}
where $\varrho_{AB}^{T_A}$ is the partially transposed matrix of $\varrho_{AB}$ with regard to the subsystem $A$, and $||Y||$ represents the trace norm of $Y$, that is, $||Y||=\mathrm{Tr}\sqrt{YY^\dag}$.
It vanishes iff $\varrho_{AB}$ is separable for $2\otimes2$ and $2\otimes3$ quantum states \cite{Horodecki52391998}. For convenience, we use the following definition of negativity,
$N(\varrho_{AB})=||\varrho_{AB}^{T_A}||-1$.

For any  $d\otimes d$ bipartite pure state $|\psi\rangle_{AB}$ with Schmidt decomposition form, $|\psi\rangle_{AB}=\sum_{i=1}^d\sqrt{\vartheta_i}|ii\rangle$,
one gets
\begin{eqnarray}\label{ne}
 N(|\psi\rangle_{AB})=2\sum_{i<j}\sqrt{\vartheta_i\vartheta_j}.
\end{eqnarray}
Based on the definition of concurrence (\ref{defconcurrence}), we obtain
\begin{eqnarray}\label{co}
 \mathcal{C}(|\psi\rangle_{AB})=2\sqrt{\sum_{i<j}\vartheta_i\vartheta_j}.
\end{eqnarray}
From (\ref{ne}) and (\ref{co}), one has
\begin{eqnarray}\label{nac}
N(|\psi\rangle_{AB})\geq \mathcal{C}(|\psi\rangle_{AB}).
\end{eqnarray}
Particularly, for any bipartite pure state $|\psi\rangle_{AB}$ with Schmidt rank 2,
it holds that  $N(|\psi\rangle_{AB})=\mathcal{C}(|\psi\rangle_{AB})$.

For a mixed state $\varrho_{AB}$, the convex-roof extended negativity  (CREN) is defined as follows
\begin{equation*}
 N_c(\varrho_{AB})=\mathrm{min}\sum_ip_iN(|\psi_i\rangle_{AB}),
\end{equation*}
where the minimum is taken over all possible pure state decompositions $\{p_i,\,|\psi_i\rangle_{AB}\}$ of $\varrho_{AB}$. CREN provides a perfect means of discrimination between positively partial transposed bound entangled states and separable states in any bipartite quantum systems \cite{Horodeki3331997,D0623132000}. CREN is equivalent to concurrence for any pure state with Schmidt rank 2 \cite{Kim0123292009}.
As a consequence, for any two-qubit mixed state $\varrho_{AB}$, one gets
\begin{eqnarray}\label{N1}
 \mathcal{C}(\varrho_{AB})=N_c(\varrho_{AB}).
\end{eqnarray}

For an $N$-qubit state $\varrho_{AB_1\cdots B_{N-1}}\in H_A\otimes H_{B_1}\otimes\cdots\otimes H_{B_{N-1}}$, it has been shown that \cite{Kim0123292009}
\begin{eqnarray}\label{sum1n}
&&N^\nu_c(\varrho_{A|B_1B_2\cdots B_{N-1}})\nonumber\\
 &&\geq N^\nu_c(\varrho_{AB_1})+ N^\nu_c(\varrho_{AB_2})+\cdots+N^\nu_c(\varrho_{AB_{N-1}})
\end{eqnarray}
for $\nu\geq 2$.
Inequality (\ref{sum1n}) is a summation-form monogamy relation based on CREN. Later, Zhang {\it  et. al} presented the product-form monogamy relations in terms of the $\nu$-th power of CREN~\cite{Zhang21500222021},
\begin{equation}\label{product1.1n}
N_c^\nu(\varrho_{A|BC})\geq
\big(4N_c(\varrho_{AB})^2N_c(\varrho_{AC})^2+\epsilon_{ABC}^2\big)^{\frac{\nu}{4}},
\end{equation}
where $\nu\geq2$, $\epsilon_{ABC}=N_c^2(\varrho_{A|BC})-\big(N_c^2(\varrho_{AB})+N_c^2(\varrho_{AC})\big)$.

Similar to the lemma \ref{productL1} in Sec.\ref{s1}, the following conclusion is obtained by us.

\begin{lemma}\label{productL3}
For any three-qubit pure  state $|\psi\rangle_{ABC}\in H_A\otimes H_B\otimes H_C$, we have
\begin{equation}\label{product2n}
\begin{aligned}
N^{\nu}_{cA|BC}\geq \big[4(N_{c}^2(\varrho_{AB})+\frac{\epsilon_{ABC}}{2})(N_{c}^2(\varrho_{AC})
+\frac{\epsilon_{ABC}}{2})\big]^{\frac{\nu}{4}}
\end{aligned}
\end{equation}
for $\nu\geq2$.

\begin{proof}
For any three-qubit pure state $|\psi\rangle_{ABC}$, one has \cite{Ou0623082007}
\begin{eqnarray*}
 N_c^{2}(|\psi\rangle_{ABC})=\mathcal{C}^{2}(|\psi\rangle_{ABC}).
\end{eqnarray*}
From the relation (\ref{N1}) and Lemma \ref{productL1}, we have
\begin{equation}
\begin{array}{rl}
N_{c}^2(\varrho_{A|BC})\geq 2\big[(N_{c}^2(\varrho_{AB})+\frac{\epsilon_{ABC}}{2})(N_{c}^2(\varrho_{AC})
+\frac{\epsilon_{ABC}}{2})\big]^{\frac{1}{2}}.
\end{array}
\end{equation}
Hence, for $\nu\geq2$ it follows that
\begin{equation*}
\begin{array}{rl}
&N_{c}^\nu(\varrho_{A|BC})\\
&=\big(N_{c}^2(\varrho_{A|BC})\big)^{\frac{\nu}{2}}\\[1mm]
&\geq \Big[2\sqrt{(N_{c}^2(\varrho_{AB})+\frac{\epsilon_{ABC}}{2})
(N_{c}^2(\varrho_{AC})+\frac{\epsilon_{ABC}}{2})}\Big]^{\frac{\nu}{2}}\\[1mm]
&=\big[4(N_{c}^2(\varrho_{AB})+\frac{\epsilon_{ABC}}{2})(N_{c}^2(\varrho_{AC})
+\frac{\epsilon_{ABC}}{2})\big]^{\frac{\nu}{4}}.
\end{array}
\end{equation*}
\end{proof}
\end{lemma}

The product-form monogamy inequality in Lemma 3 is tighter than the inequality (\ref{product1.1n}) presented  in Ref.~\cite{Zhang21500222021}.
Based on Lemma \ref{productL3}, we present the product-form monogamy relation for $N$-qubit states. For convenience, we denote by $N_{cAB_j}=N_c(\varrho_{AB_j})$ for $j=1,2,\cdots,N-1$, and $N_{cA|B_1B_2\cdots B_{N-1}}=N_c(\varrho_{A|B_1 B_2\cdots B_{N-1}})$. Similar to the way of proving Theorem \ref{productT1}, through the utilization of the inequality(\ref{sum1n}) and Lemma \ref{productL3} we obtain the following result.

\begin{theorem}\label{productT2}
For any $N$-qubit pure quantum state $\varrho_{AB_1\cdots B_{N-1}}\in H_{A}\otimes H_{B_1}\otimes \cdots\otimes H_{B_{N-1}}$, we have
\begin{equation}\label{product4n}
\begin{array}{rl}
N_{cA|B_1\cdots B_{N-1}}^\nu\geq
&\Big(4(N^2_{cAB_{1}}+\frac{\epsilon_{AB_{1}\cdots B_{N-1}}}{2})\\
&\big((N-2)\Big(\prod\limits_{i=1}^{N-2}N^2_{cAB_{i+1}}\Big)^{\frac{1}{N-2}}\\
&+\frac{\epsilon_{AB_{1}\cdots B_{N-1}}}{2}\big)\Big)^{\frac{\nu}{4}},
\end{array}
\end{equation}
where $\epsilon_{AB_1\cdots B_{N-1}}=N^2_{cA|B_1\cdots B_{N-1}}-\big(N^2_{cAB_1}+N^2_{cAB_2\cdots B_{N-1}}\big).$
\end{theorem}

We take into account the following example to show that our lower bound (\ref{product4n}) is tighter than the one given in the summation-form monogamy relation (\ref{sum1n}) \cite{Kim0123292009}.

\begin{figure}[htbp]
\centering
{\includegraphics[width=6cm,height=4.5cm]{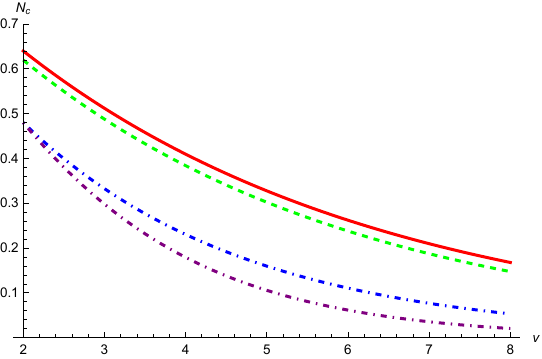}}
\caption{Solid red line denotes $N^\nu_{cA|BC}$ for the state given in Eq.(\ref{GSD}). The green thick dotted (blue dot dashed thick, purple dot dashed thick) line represents the lower bound from our result (\ref{product2n}) (Eq.(\ref{product1.1n}) in Ref.~\cite{Zhang21500222021} and Eq.(\ref{sum1n}) in Ref.~\cite{Kim0123292009}, respectively).} \label{Fig2}
\end{figure}

{\noindent\bf Example 2}.
Let us consider the three-qubit state $|\phi\rangle_{ABC}$ in the generalized Schmidt decomposition~\cite{Tr39},
\begin{equation}\label{GSD}
|\phi\rangle_{ABC}=\vartheta_0|000\rangle+\vartheta_1e^{i\varphi}|100\rangle+\vartheta_2|101\rangle+\vartheta_3|110\rangle+\vartheta_4|111\rangle,
\end{equation}
where $\vartheta_i\geq0$, $i=0,1,\cdots,4$, and $\sum\limits_{i=0}^{4}\vartheta_i^2=1$.
One gets $N_{cA|BC}=2\vartheta_0\sqrt{\vartheta_2^2+\vartheta_3^2+\vartheta_4^2}$, $N_{cAB}=2\vartheta_0\vartheta_2$ and $N_{cAC}=2\vartheta_0\vartheta_3$. Setting $\vartheta_0=\vartheta_3=\vartheta_4=\sqrt{\frac{1}{5}}$, $\vartheta_2=\sqrt{\frac{2}{5}}$ and $\vartheta_1=0$, we have $N_{cA|BC}=\frac{4}{5}$, $N_{cAB}=\frac{2\sqrt{2}}{5}$, $N_{cAC}=\frac{2}{5}$ and $\epsilon_{ABC}=\frac{4}{25}$. Then
$\big(4N_{cAB}^2N_{cAC}^2+\epsilon_{ABC}^2\big)^{\frac{\nu}{4}}
=\big(4(\frac{2\sqrt{2}}{5})^2(\frac{2}{5})^2+(\frac{4}{25})^2\big)^{\frac{\nu}{4}}$ from Eq.(\ref{product1.1n}) in Ref.~\cite{Zhang21500222021}, $N^{\nu}_{cAB}+N^{\nu}_{cAC}
=(\frac{2\sqrt{2}}{5})^{\nu}+(\frac{2}{5})^{\nu}$
from Eq.(\ref{sum1n}) in Ref.~\cite{Kim0123292009} and $\big[4(N^2_{cAB}+\frac{\epsilon_{ABC}}{2})(N^2_{cAC}+\frac{\epsilon_{ABC}}{2})\big]^{\frac{\nu}{4}}
=\big[4((\frac{2\sqrt{2}}{5})^2+\frac{2}{25})((\frac{2}{5})^2+\frac{2}{25}))\big]^{\frac{\nu}{4}}$
from our result (\ref{product2n}). It is evident that our result (\ref{product2n}) is better than the results Eq.(\ref{product1.1n}) and Eq.(\ref{sum1n}) given in Ref.~\cite{Zhang21500222021} and Ref.~\cite{Kim0123292009}, respectively, see Fig.\ref{Fig2}.

CREN can be regarded as a generalized form of concurrence from 2-qubit systems. Therefore, with the monotonicity and separability criteria of CREN in place, it is natural to explore the MOE in terms of CREN for higher-dimensional quantum systems. Here we show that our result (\ref{product2n}) still holds for the two counterexamples given in Refs.~\cite{ou, ks}, while the CKW inequality in terms of concurrence is violated.
\begin{counterexample}\label{Ex:1} (Ou~\cite{ou}) Consider the following $3 \otimes 3\otimes 3$ pure state,
\small\begin{align}  \label{f}
|\psi\rangle_{ABC}=\frac{1}{\sqrt{6}}(&|123\rangle-|132\rangle+|231\rangle
-|213\rangle+|312\rangle-|321\rangle).
\end{align}\normalsize
We have ${N}_{cA|BC}=2$, ${N}_{cAB}={N}_{cAC}=1$ and $\epsilon_{ABC}=2$. Therefore,
\begin{equation}
\begin{aligned}
N^{\nu}_{cA|BC}=2^{\nu}\geq 2^{\nu}= \big[4(N^2_{cAB}+\frac{\epsilon_{ABC}}{2})(N^2_{cAC}+\frac{\epsilon_{ABC}}{2})\big]^{\frac{\nu}{4}}.\nonumber
\end{aligned}
\end{equation}
\end{counterexample}

\begin{counterexample}\label{EX2}
(Kim and Sanders~\cite{ks})
Consider the $3 \otimes 2 \otimes 2$ pure state $\ket{\varphi}$,
\begin{equation}
\ket{\varphi}_{ABC} = \frac{1}{\sqrt{6}}(\sqrt{2}\ket{010}+\sqrt{2}\ket{101}+\ket{200}+\ket{211}).
\label{count2}
\end{equation}
It is verified that ${N}^{2}_{cA|BC}=4$, ${N}^{2}_{cAB}={N}^{2}_{cAC}=\frac{8}{9}$ and $\epsilon_{ABC}=\frac{20}{9}$. We have
\begin{equation}
\begin{aligned}
N^{\nu}_{cA|BC}=2^{\nu}\geq 2^{\nu}= \big[4(N^2_{cAB}+\frac{\epsilon_{ABC}}{2})(N^2_{cAC}+\frac{\epsilon_{ABC}}{2})\big]^{\frac{\nu}{4}}.\nonumber
\end{aligned}
\end{equation}
\end{counterexample}

Although the states (\ref{f}) and (\ref{count2}) are two counterexamples of the CKW inequality in terms of concurrence, they still satisfy our product-form monogamy
inequality (\ref{product2n}).

\section{CONCLUSION}\label{s3} We have presented tighter monogamy inequalities in product-form by using the concurrence and negativity. Compared with the existing monogamy relations, our product-form monogamy relations of multi-qubit quantum entanglement have tighter lower bounds. Moreover, our product-form monogamy relation in terms of CREN is still valid for the counterexamples for which the CKW inequality is violated. Our tighter product-form monogamy inequalities lead to finer characterization of entanglement distribution among subsystems. Our approach may also highlight further investigations on the sharability of other quantum correlations.\\

\bigskip
\section*{ACKNOWLEDGMENTS}
This work is supported by the National Natural Science Foundation of China under Grant Nos. 12075159 and 12171044; the specific research fund of the Innovation Platform for Academicians of Hainan Province.

\bigskip
\noindent{\bf Data availability statement:}
 All data generated or analyzed during this study are included and cited in this article.

\bigskip
\noindent{\bf Conflicts of Interest:} The authors declare no conflict of interest.

\end{document}